\documentstyle[11pt]{article}

\newcommand{\be}{\begin{equation}}
\newcommand{\ee}{\end{equation}}
\newcommand{\bea}{\begin{eqnarray}}
\newcommand{\eea}{\end{eqnarray}}

\begin{document}
\begin{titlepage}

\vspace{0.2in}

\begin{flushright}

OUTP-92-42P

CPT-92/PE.2840

December 1992

\end{flushright}

\vspace*{0.5cm}
\begin{center}

{\bf THE SHIFTED COUPLED CLUSTER METHOD: A NEW APPROACH TO
HAMILTONIAN LATTICE GAUGE THEORIES \\}
\vspace*{1cm}
{\bf C H Llewellyn Smith} \\
Department of Physics, Theoretical Physics, 1 Keble Road, \\
Oxford OX1 3NP, England. \\
\vspace*{0.5cm}
{\bf N J Watson} \\
Centre de Physique Th\'eorique, C.N.R.S. - Luminy, Case 907, \\
F-13288 Marseille Cedex 9, France. \\

\vspace*{1.5cm}
{\bf   Abstract  \\ }
\end{center}
It is shown how to adapt the non-perturbative coupled cluster method
of many-body theory so that it may be successfully applied to Hamiltonian
lattice $SU(N)$ gauge theories.
The procedure involves first writing the wavefunctions for the vacuum
and excited states in terms of linked clusters of gauge invariant
excitations of the strong coupling vacuum.
The fundamental approximation scheme then consists of
i) a truncation of the infinite set of clusters in the wavefunctions
according to their geometric {\em size}, with all larger clusters
appearing in the Schr\"odinger equations simply discarded,
ii) an expansion of the truncated wavefunctions in terms of the
remaining clusters rearranged, or ``shifted'', to describe gauge
invariant {\em fluctuations} about their vacuum expectation values.
The resulting non-linear truncated Schr\"odinger equations are then solved
self-consistently and exactly.
Results are presented for the case of $SU(2)$ in $d=3$ space-time dimensions.
\vfill
\end{titlepage}

This letter reports the first results of a new approach to
Hamiltonian lattice gauge theories. The goal is to find approximate
non-perturbative solutions of the Schr\"odinger equation
for both vacuum and excited states. Space is treated as a discrete lattice
in order to  i) reduce field theory to $n \rightarrow\infty$
body quantum mechanics, ii) render the
theory ultra-violet finite, and  iii) make trivial the imposition of Gauss's
law, which must be
imposed as a subsidiary condition on the states in an $A_{0} = 0$ gauge,
which is the natural choice with a Hamiltonian
approach\footnote{
To avoid ghosts, a non-covariant gauge is necessary, which must be
$A_{0} = 0$ to retain rotational invariance. In contrast to QED,
Gauss's law cannot be solved explicitly \cite{gribov}.}.

The basic variables are the $SU(N)$ matrices
$U(L)$ defined on each link $L$. Conjugate
``chromoelectric'' fields $E^{a}(L)$, $a = 1\ldots N^{2}-1$, are given
by\footnote{
This operation, in which the elements of the unitary matrices
$U(L)$ are treated as if they were independent,
and with the indices $i,j$ summed over, correctly reproduces the
fundamental commutation relations for $E^{a}(L)$ with $U(L')$,
$U^{\dagger}(L')$ and $E^{a'}(L')$.}
\be
E^{a}(L)
=
-\Bigl(T^{a}U(L)\Bigr)_{ij}\frac{\partial}{\partial U(L)_{ij}}
+\Bigl(U^{\dagger}(L)T^{a}\Bigr)_{ij}
\frac{\partial}{\partial U^{\dagger}(L)_{ij}}
\ee
in Schr\"odinger representation, where $T^{a}$ are the group generators.
The Hamiltonian for $SU(N)$ in $d$ space-time
dimensions, without quarks, may be written \cite{kogsuss}
\be
H(x)
=
a^{-1}x^{-1/2}W(x)
\ee
where the dimensionless variable $x$ is related to the coupling
$g$ by $x = 4g^{-4}a^{2d-8}$, and
\be
W(x)
=
\sum_{L}E^{a}(L)^{2}
- x\sum_{p}{\rm Re}\,U_{p}
\ee
where $U_{p}$ is the trace of the product of
$U$s  and  $U^{\dagger}$s around the four sides of
a plaquette $p$:
$U_{p} = {\rm Tr}U(1)U(2)U^{\dagger}(3)U^{\dagger}(4)$.
The Schr\"odinger equation for the ``reduced'' energy  $w = ax^{1/2}E$ is then
\be
W\Psi
=
w\Psi
\ee
The continuum limit occurs at $x\rightarrow \infty$.

We consider first perturbation theory about the strong coupling $x = 0$
vacuum, but emphasize that the method we shall develop applies at {\em all}
couplings $0 \leq x < x_{c}$, where $x_{c}$ is the lowest $x$ at which there
is a phase transition.
For $SU(N)$, it is expected that $x_{c} = \infty$ i.e. the continuum limit.
At $x \ll 1$, the vacuum wavefunction $\Psi_{0}$ of, for example,
the $SU(2)$ theory in $d = 3$ is

\addtocounter{equation}{1}
\noindent
\begin{picture}(425,125)(0,20)

\put(20,120){\makebox(0,0)[l]{${\displaystyle
\Psi_{0}
=
1 + x\biggl({\textstyle \frac{1}{3}}\sum_{s}\hspace{15pt}\biggr)
}$}}

\put(105,116){\framebox(10,10){}}

\put(47,80){\makebox(0,0)[l]{${\displaystyle
+ \hspace{3pt} x^{2}\biggl({\textstyle\frac{1}{18}}
\sum_{s}\hspace{15pt}\sum_{s'}\hspace{15pt}
-{\textstyle \frac{1}{72}}\sum_{s}\hspace{15pt}
-{\textstyle \frac{1}{117}}\sum_{s}\Bigl(\hspace{23pt}+\hspace{15pt}\Bigr)
+{\textstyle \frac{8}{351}}\sum_{s}\Bigl(\hspace{23pt}+\hspace{15pt}\Bigr)
\biggr)
}$}}

\put(106,77){\framebox(10,10){}}
\put(137,77){\framebox(10,10){}}
\put(190,77){\framebox(10,10){}}
\put(192,79){\framebox(6,6){}}
\put(252,77){\framebox(9,10){}}
\put(263,77){\framebox(9,10){}}
\put(284,77){\framebox(10,9){}}
\put(284,88){\framebox(10,9){}}
\put(355,77){\framebox(20,10){}}
\put(387,77){\framebox(10,20){}}

\put(47,40){\makebox(0,0)[l]{${\displaystyle
+ \hspace{3pt} x^{3}\biggl({\textstyle \frac{1}{162}}
\sum_{s}\hspace{15pt}\sum_{s'}\hspace{15pt}\sum_{s''}\hspace{15pt}
-{\textstyle \frac{1}{216}}
\sum_{s}\hspace{15pt}\sum_{s'}\hspace{15pt}
+\ldots
-{\textstyle \frac{881}{63180}}\sum_{s}\hspace{15pt}
\biggr) +\ldots \hspace{10pt}(5)
}$}}

\put(110,37){\framebox(10,10){}}
\put(141,37){\framebox(10,10){}}
\put(172,37){\framebox(10,10){}}
\put(230,37){\framebox(10,10){}}
\put(232,39){\framebox(6,6){}}
\put(261,37){\framebox(10,10){}}
\put(354,37){\framebox(10,10){}}

\end{picture}

\noindent
where the sums are over all sites $s$ on the lattice.
As shown originally by Hubbard \cite{hub}
in the case of continuum field theory,
and observed by Greensite for lattice theories \cite{green},
the products of sums of independent excitations exponentiate:
\be
\Psi_{0}
=
{\rm exp}S
\ee
For {\em all} couplings $0 \leq x < x_{c}$, and for
general $SU(N)$ and $d$, the exponential form (6) holds, where
the function $S$ is a single sum over the sites $s$ of the lattice
of all possible linearly independent ``linked clusters'' $C_{i,s}$
of $SU(N)$ Wilson loops:
\be
S
=
\sum_{s}\sum_{i}a_{i}C_{i,s}
\ee
By ``linked'' is meant products
of Wilson loops at {\em fixed} relative separation
and orientation. The ``amplitudes'' $a_{i}$ are then the coefficients of
these clusters, summed to all orders in strong coupling perturbation theory.
The representation (6),(7) holds, and is
exact, provided the true vacuum is not orthogonal to the strong coupling
vacuum $\Psi_{0}(x\!=\!0)$ i.e. up to the lowest $x$ at which there is
a phase transition. Thus, for any  $N$ and $d$, the Schr\"odinger
equation for the vacuum energy is equivalent to
\be
\sum_{L}\Bigl(E^{a}(L)^{2}S + E^{a}(L)S \cdot E^{a}(L)S \Bigr)
- x\sum_{p}{\rm Re}\,U_{p}
=
\sum_{s}w_{0}
\ee

The wavefunctions of excited states (glueballs) with zero
momentum can be written
\be
\Psi_{g}
=
G\,{\rm exp}S
\ee
where
\be
G
=
\sum_{s}\Bigl(b_{0} + \sum_{i}b_{i}C_{i,s}\Bigr)
\ee
i.e. $G$ is a similar
sum\footnote{
Excited states of a given spin and parity are constructed through the
appropriate dependence of the $b_{i}$ on the clusters' rotation/reflection
orientations, left implicit in (10).}
over the lattice of linked clusters of $SU(N)$
Wilson loops. The corresponding Schr\"odinger equation is then
\be
\sum_{L}\Bigl( E^{a}(L)^{2}G + 2E^{a}(L)S \cdot E^{a}(L)G \Bigr)
=
(w_{g} - w_{0})G
\ee

The effect of the electric field operators
on a given link $L$ in (8), (11) is to recombine
the clusters in $S$ and $G$ containing $U(L)$ or $U^{\dagger}(L)$
according to the relation for $SU(N)$ generators
$ T_{\alpha\beta}^{a}T_{\gamma\delta}^{a} =
\frac{1}{2}\delta_{\alpha\delta}\delta_{\beta\gamma}
-\frac{1}{2N}\delta_{\alpha\beta}\delta_{\gamma\delta} $.
Some examples for general $SU(N)$ are

\noindent
\begin{picture}(425,50)(0,5)

\put(30,30){\makebox(0,0)[l]{${\displaystyle
E^{a}(L)^{2}\hspace{35pt} = {\textstyle \frac{N^{2}-1}{2N}}\hspace{60pt}
}$}}

\put(70,25){\framebox(20,10){}}
\put(70,25){\circle*{3}}
\put(80,25){\circle*{3}}
\put(90,25){\vector(0,1){8}}
\put(145,25){\framebox(20,10){}}
\put(165,25){\vector(0,1){8}}

\put(200,30){\makebox(0,0)[l]{${\displaystyle
E^{a}(L)\hspace{15pt}\cdot E^{a}(L)\hspace{15pt} =
{\textstyle \frac{1}{2}}\hspace{20pt}
-{\textstyle \frac{1}{2N}}\hspace{20pt}
}$}}

\put(233,30){\framebox(10,10){}}
\put(233,30){\circle*{3}}
\put(243,30){\circle*{3}}
\put(243,30){\vector(0,1){8}}

\put(285,20){\framebox(10,10){}}
\put(285,30){\circle*{3}}
\put(295,30){\circle*{3}}
\put(295,30){\vector(0,-1){8}}

\put(322,20){\line(1,0){10}}
\put(322,20){\line(0,1){8}}
\put(322,28){\line(5,2){10}}
\put(322,32){\line(0,1){8}}
\put(322,32){\line(5,-2){10}}
\put(322,40){\line(1,0){10}}
\put(332,20){\line(0,1){8}}
\put(332,32){\line(0,1){8}}
\put(332,32){\vector(0,1){6}}
\put(332,28){\vector(0,-1){6}}

\put(370,20){\framebox(10,9){}}
\put(370,31){\framebox(10,9){}}
\put(380,31){\vector(0,1){7}}
\put(380,29){\vector(0,-1){7}}

\end{picture}

\noindent
where the dots indicate the ends of the link $L$.

The exponential form (6) for the vacuum wavefunction
is in fact a very general expression
for the fully interacting ground state of a many-body system, with
$S$ a sum of linked clusters each of which describes linked
excitations of a bare or unperturbed state $\Psi_{0}(x\!=\!0)$.
The exponentiation
then gives the correct statistical weighting for unlinked excitations
consisting of products of linked components.
Furthermore, it guarantees that the Schr\"odinger
equation for the vacuum and for excited states
is a {\em single} sum of linked terms only, so that there are then
no size extensivity problems with approximation schemes.
The expression (6) is
the starting point for the many-body theory coupled cluster
method\footnote{For an
introductory review of this method, which has been widely used
in nuclear and
condensed matter physics and quantum chemistry, see ref. \cite{bish}.
Application to lattice gauge theory was suggested by
Greensite \cite{green}, but this suggestion was not followed up.}.
This is an intrinsically non-perturbative approach, involving
the amplitudes $a_{i}, b_{i}$
directly rather than their perturbative expansions.

In order to apply the coupled cluster method to lattice $SU(N)$
gauge theories,
the fundamental approximation in general consists of two steps:

I) A truncation of the sums of clusters in $S$ and $G$
according to the clusters' geometrical {\em size}. All clusters then formed
in the Schr\"odinger equations with the truncated $S$ and $G$
which are larger than the given cut-off are simply discarded.

II) A Taylor expansion of $S$ and $G$ in the remaining clusters, rearranged,
or ``shifted'',
to describe {\em fluctuations} of the individual Wilson loops about
their vacuum expectation values. These
expectation values are calculated self-consistently from the vacuum
wavefunction.

The first step was originally suggested by Greensite \cite{green}, but
in order for the method to provide a tractable,
valid and efficient approximation scheme beyond
the perturbative range of couplings, the second step is in general essential.
The physical idea is that amplitudes in $S$ involving
fluctuations on scales larger than the vacuum correlation length $\xi$,
and in $G$ larger than the size $\rho$ of the glueball,
are expected to be greatly reduced relative to those for fluctuations
smaller than $\xi$ and $\rho$, and to decrease very rapidly with
increasing size.
For fixed lattice spacing, the results should therefore converge rapidly
to the exact result as the dimensions of the cut-off cluster size are increased
beyond $\xi$ and $\rho$. As the lattice spacing $a \rightarrow 0$, $\xi/a$
and $\rho/a$ diverge and the size in lattice units of clusters required to
describe the theory must diverge correspondingly. The success of the
method therefore depends on the results obtained at
successsively increasing values of $x$ (i.e. decreasing $a$)
showing a sufficiently rapid approach to the expected scaling behaviour
that the continuum limit can be extrapolated.

We consider first the case of $SU(2)$ with a one-plaquette cut-off,
for which step I is exactly soluble. For $SU(2)$, the identity
${\rm Tr}A\,{\rm Tr}B = {\rm Tr}AB + {\rm Tr}AB^{-1}$ for $SU(2)$ matrices
$A$, $B$ gives linear relations between clusters e.g.

\noindent
\begin{picture}(425,50)(0,5)

\put(50,20){\framebox(14,15){}}
\put(66,20){\framebox(29,15){}}
\put(105,27.5){\makebox(0,0){${\displaystyle = }$}}
\put(115,20){\framebox(45,15){}}
\put(170,27.5){\makebox(0,0){${\displaystyle + }$}}
\put(180,20){\line(1,0){13.5}}
\put(180,20){\line(0,1){15}}
\put(180,35){\line(1,0){13.5}}
\put(193.5,20){\line(1,5){3}}
\put(193.5,35){\line(1,-5){3}}
\put(196.5,20){\line(1,0){28.5}}
\put(196.5,35){\line(1,0){28.5}}
\put(225,20){\line(0,1){15}}

\put(290,20){\framebox(15,15){}}
\put(292.5,22.5){\framebox(10,10){}}
\put(315,27.5){\makebox(0,0){${\displaystyle = }$}}
\put(330,27.5){\makebox(0,0){${\displaystyle 2 }$}}
\put(345,27.5){\makebox(0,0){${\displaystyle + }$}}
\put(355,20){\line(1,0){15}}
\put(355,20){\line(0,1){15}}
\put(355,35){\line(1,0){15}}
\put(357.5,22.5){\line(1,0){10}}
\put(357.5,22.5){\line(0,1){10}}
\put(357.5,32.5){\line(1,0){10}}
\put(367.5,22.5){\line(1,5){2.5}}
\put(367.5,32.5){\line(1,-5){2.5}}

\end{picture}

\noindent
Thus, the one-plaquette clusters may be written as ``powers'' of the
one-plaquette single loop $C_{1,p}$. Then, with the one-plaquette
cut-off $S$, writing the infinite sum of one-plaquette
clusters on each $p$ as a function $S_{p}(C_{1,p})$
\be
S
=
\sum_{p}S_{p}(C_{1,p})
\ee
Schr\"odinger's equation for the vacuum becomes the differential equation
\be
\sum_{p}\Biggl\{
\frac{dS_{p}}{dC_{1,p}}3C_{1,p} +
\Biggl(\frac{d^{2}S_{p}}{dC_{1,p}^{2}} +
\biggl(\frac{dS_{p}}{dC_{1,p}}\biggr)^{2}\Biggr)\Biggl(C_{1,p}^{2} - 4 \Biggr)
-xC_{1,p}\Biggr\}
=
\sum_{s}w_{0}
\ee
where terms involving clusters spanning adjacent plaquettes $p,p'$
\be
\frac{dS_{p}}{dC_{1,p}}\frac{dS_{p'}}{dC_{1,p'}}
E^{a}C_{1,p}\cdot E^{a}C_{1,p'}
\ee
have been discarded. The solution is\footnote{
Eqn. (15), which is the exact solution for an $SU(2)$
``one plaquette universe'', has been
derived by many authors e.g. refs. \cite{kato}, \cite{robweb},
and also as the $d = 3$
variational solution if $\Psi$ is taken to be a product of functions
of one plaquette variables \cite{duncan}.
However, the fact that it reproduces
so well the results of ref. \cite{hamirv3d}
 for $d=3$ has not been noticed previously.
For $d=4$ (for which a variational solution is not
available) eqn. (15) still results in a vacuum energy
which compares well with that obtained from Pad\'e approximants
derived from strong coupling series \cite{hamirv4d}.}
\be
S_{p}
=
{\rm ln}\,\Biggl(\Bigl(1 - {\textstyle \frac{1}{4}}C_{1,p}^{2} \Bigr)^{-1/2}
se_{2}\biggl(-{\textstyle \frac{1}{2}}{\rm cos}^{-1}
\Bigl(-{\textstyle \frac{1}{2}}C_{1,p}\Bigr)\biggr)\Biggr)
\ee
where $se_{2}$ is a Mathieu function \cite{abrsteg}.
The associated vacuum energy per site $w_{0}$ for $d=3$ is plotted
in fig. 1, together with Pad\'e approximants obtained from the strong
coupling series to ${\cal O}(x^{14})$ \cite{hamirv3d}
with which it shows very remarkable agreement at all $x$.
Furthermore, using the Feynman-Hellman theorem,
the vacuum expectation value
$\langle C_{1}\rangle = -dw_{0}/dx$,
has the correct weak coupling behaviour
\be
\hspace{1cm} \langle C_{1}\rangle
=
2 - {\cal O}(x^{-1/2}) \hspace{1cm} x \gg 1
\ee

The first excited state involves the Mathieu function $se_{4}$. The
mass gap $\Delta w_{g} = w_{g} - w_{0}$ for $d = 3$
is shown plotted against $x^{-1/2}$
in fig. 2, together with Pad\'e approximants
from ${\cal O}(x^{10})$ series
\cite{hamirv3d}.
In $d = 3$, the $SU(2)$ theory is superrenormalizable and so the true
(reduced) mass gap $\Delta w_{g}$ is expected to go to a constant at the
continuum limit $x \rightarrow \infty$.
At weak coupling, the Mathieu mass gap diverges as
$\Delta w_{g} = 4x^{1/2} - \frac{5}{4} + {\cal O}(x^{-1/2})$. This $x^{1/2}$
behaviour is as expected from the uncertainty principle with a wavefunction
involving excitations spanning only a finite number of lattice spacings.

In general, increasing the cut-off size beyond one plaquette
or going beyond $SU(2)$, the
resulting partial differential equation for $S_{p}$ is intractable.
We therefore introduce the second step of the approximation scheme and
rearrange the sums of clusters in
the given cut-off $S$ and $G$ in terms of ``shifted'' linked clusters
$C_{i,s}'$
\bea
S
&=&
\sum_{s}\sum_{i}a_{i}'C_{i,s}' \\
G
&=&
\sum_{s}\Bigl( b_{0}' + \sum_{i}b_{i}'C_{i,s}' \Bigr)
\eea
These shifted linked clusters
consist of products of Wilson loops each minus their vacuum
expectation values, still at fixed relative separation and
orientation e.g. for general $SU(N)$

\noindent
\begin{picture}(425,45)(0,5)

\put(40,25){\makebox(0,0)[l]{${\displaystyle
\Bigl( \hspace{30pt} \Bigr)'
}$}}

\put(50,20){\framebox(20,10){}}
\put(60,20){\vector(1,0){8}}
\put(52,22){\framebox(7,6){}}
\put(59,28){\vector(0,-1){6}}

\put(97,25){\makebox(0,0)[l]{
{\rm with}
}}

\put(140,20){\framebox(20,10){}}
\put(150,20){\vector(1,0){8}}
\put(162,28){\makebox(0,0)[l]{${\displaystyle
' \hspace{6pt} =
}$}}
\put(190,20){\framebox(20,10){}}
\put(200,20){\vector(1,0){8}}
\put(220,25){\makebox(0,0)[l]{${\displaystyle
- \hspace{8pt}\langle\hspace{25pt}\rangle,
}$}}
\put(243,20){\framebox(20,10){}}
\put(253,20){\vector(1,0){8}}

\put(300,20){\framebox(10,10){}}
\put(310,30){\vector(0,-1){8}}
\put(312,28){\makebox(0,0)[l]{${\displaystyle
' \hspace{6pt} =
}$}}
\put(340,20){\framebox(10,10){}}
\put(350,30){\vector(0,-1){8}}
\put(360,25){\makebox(0,0)[l]{${\displaystyle
- \hspace{8pt}\langle\hspace{15pt}\rangle
}$}}
\put(383,20){\framebox(10,10){}}
\put(393,30){\vector(0,-1){8}}

\end{picture}

To illustrate the method, we consider again the one-plaquette cut-off and
include in $S$ and $G$ shifted one-plaquette
clusters up to the $n$'th power of the shifted one-plaquette loop
$C_{1,p}' = C_{1,p} - \langle C_{1}\rangle$.
Substituting into (8) and (11), the resulting
clusters spanning two plaquettes are discarded (step I) together with all
``higher order'' fluctuations $(C_{1,p}')^{m}$, $m > n$, (step II)
and coefficients of the constant term and each
shifted cluster $(C_{1,p}')^{i}$, $i = 1$ to $n$, are equated.
For the vacuum, this gives a set of $n$ non-linear
equations for the amplitudes $a_{i}'$ and an expression
for $w_{0}$, all in terms of $x$ and $\langle C_{1}\rangle$.
For the glueball, the procedure gives an $n \times n$ matrix equation
for the $b_{i}'$ with eigenvalues $\Delta w_{g}$,
together with an equation for $b_{0}'$,
all in terms of the $a_{i}'$ and $\langle C_{1} \rangle$.
These coupled cluster equations are then solved
using numerical library routines, with
$\langle C_{1} \rangle$ calculated self-consistently

The results for $w_{0}$ and $\Delta w_{g}$ obtained
\cite{thesis} with the one plaquette cut-off and
the expansions of $S$ and $G$ truncated at  $n = 1,2,3\ldots$
converge rapidly to the Mathieu results at all $x$.
Each successive power of $C_{1,p}'$ generates the next term in
the series expansions in $x$ for $x \ll 1$ and $x^{-1/2}$ for $x \gg 1$
of the Mathieu eigenvectors and eigenvalues. In particular,
even at the very simplest $n=1$ approximation
the method gives the correct {\em weak} coupling
leading behaviour (16) for the
vacuum expectation value $\langle C_{1} \rangle$.

Proceeding to larger cut-offs with $SU(2)$ in $d = 3$,
we have included in $S$ and $G$
all the shifted clusters which, unshifted, occur in
2nd, 3rd and 4th order strong coupling perturbation theory (which is
used solely as a guide to which terms to include - the
resulting coupled cluster equations are all solved non-perturbatively).
The complete set

\noindent
\begin{picture}(425,330)(0,-25)

\put(215,170){\makebox(0,0)[b]{
\begin{tabular}{cccccc}\hline
 $x$ &{\rm order}\,1&{\rm order}\,2&{\rm order}\,3&{\rm order}\,4&
Pad\'e \cite{hamirv3d}\\ \hline
0.5  &  -0.106298 & -0.0843647 & -0.0827613 & -0.0822829 & -0.082156 \\
1.0  &  -0.386012 & -0.332963  & -0.322072  & -0.319032 & -0.31587     \\
1.5  &  -0.779896 & -0.707363  & -0.685324  & -0.681253 & -0.67115     \\
2.0  &  -1.25000  & -1.16553   & -1.13371   & -1.13132  & -1.1162\,(2) \\
2.5  &  -1.77435  & -1.68175   & -1.64112   & -1.64197  & -1.626       \\
3.0  &  -2.33949  & -2.24077   & -2.19185   & -2.19667  & -2.183\,(3)  \\
3.5  &  -2.93665  & -2.83296   & -2.77600   & -2.78509  & -2.77        \\
4.0  &  -3.55974  & -3.45180   & -3.38692   & -3.40037  & -3.40\,(2)   \\
4.5  &  -4.20439  & -4.09265   & -4.01992   & -4.03771  & - \\
5.0  &  -4.86730  & -4.75207   & -4.67154   & -4.69358  & - \\
10.0 &  -12.1254  & -11.9822   & -11.8255   & -11.8828  & - \\
20.0 &  -28.2197  & -28.0321   & -27.7362   & -27.8358  & - \\ \hline
\end{tabular}
}}

\put(0,25){\makebox(0,0)[l]{\footnotesize{
Table 1. Vacuum energies $E_{0}$ for the first four orders of
approximation in the shifted coupled cluster
}}}
\put(0,10){\makebox(0,0)[l]{\footnotesize{
method, together with the results of Pad\'e approximants derived
from the ${\cal O}(x^{14})$ strong coupling series
}}}
\put(0,-5){\makebox(0,0)[l]{\footnotesize{
taken from \cite{hamirv3d} (estimated errors in the last figure are
given in brackets).
}}}

\end{picture}

\noindent
of 69 linearly independent clusters
occuring at 4th order are shown in the appendix.
The loop expectation values are calculated
self-consistently from the wavefunction using the Feynman-Hellman theorem
by making the change
\be
W
\rightarrow
W + \sum_{s}\sum_{{\rm loops} \in S,G}\epsilon_{i}C_{i,s}
\ee
so that the $a_{i}'$ become functions of the $\epsilon_{i}$ and $x$. Then
\be
\langle C_{i}\rangle
=
\left.\frac{\partial w_{0}}{\partial\epsilon_{i}}\right|_{\{\epsilon_{i}=0\}}
\ee
These are then simple quantities to calculate numerically, avoiding any group
integrals.

The results,
obtained using a computer to carry out all the calculations \cite{thesis},
are given in table 1
for $w_{0}$ and shown plotted in fig. 3  for $\Delta w_{g}$.
Assuming that the Pad\'e approximants are a good guide to
the exact results, the method gives excellent results for the vacuum
energy $w_{0}$ at strong and intermediate couplings
and the correct form $w_{0} = -2x + {\cal O}(x^{1/2})$ at
weak coupling. All the Wilson loop vacuum expectation values have the
correct weak coupling form, as in (16).
For the mass gap $\Delta w_{g}$, reasonable
results are obtained at intermediate couplings,
before the inevitable $x^{1/2}$ scaling sets in at weak coupling.
In addition to the eigenvalues, the method also provides
the wavefunctions
and the cluster vacuum expectation values at all couplings.
The amplitudes $a_{i}$, $b_{i}$ of the most dominant of the 69
shifted clusters
occuring in the wavefunctions at 4th order are shown in figs. 4 and 5
respectively.
At weak coupling all the amplitudes behave as
$a_{i}' = {a_{i}'}^{(1/2)}x^{1/2} + {\cal O}({\rm const.})$,
$b_{i}' = {b_{i}'}^{(0)} + {\cal O}(x^{-1/2})$,
with, for the 69 clusters occuring at 4th order,
the ${a_{i}'}^{(1/2)}$ and ${b_{i}'}^{(0)}$
ranging over six orders of magnitude.
For the vacuum, it is found that, at weak coupling,
simple shifted clusters spanning few plaquettes
and/or containing few Wilson loops dominate
over those for complicated multi-loop clusters.
This can be understood as a combinatorical effect in the
coupled cluster equations.
For the lowest (scalar) glueball, shifted clusters consisting of single
Wilson loops dominate at intermediate and weak coupling.
With the glueball calculation involving
the diagonalization of a (large) matrix, this effect is
harder to understand.
No evidence is found of a phase transition at finite $x$.

Returning to general $SU(N)$, it can be shown \cite{inprep} that
the shifting procedure
guarantees that, for any $N$ and $d$,
the approximation scheme gives the correct weak coupling
vacuum expectation value $N + {\cal O}(x^{-1/2})$ for a Wilson loop in $S$.
Thus, at $x \rightarrow \infty$, the effect of the electric field
operators coupling together e.g. a pair of shifted single loops becomes

\noindent
\begin{picture}(425,50)(175,5)

\put(200,30){\makebox(0,0)[l]{${\displaystyle
E^{a}(L)\hspace{15pt}\cdot E^{a}(L)\hspace{20pt} =
{\textstyle \frac{1}{2}}
\biggl( \hspace{20pt} - \hspace{20pt} - \hspace{20pt} \biggr)
-{\textstyle \frac{1}{2N}} \hspace{30pt} +
{\rm const.}\,{\rm term}\,{\cal O}(x^{-1/2})
}$}}

\put(233,30){\framebox(10,10){}}
\put(233,30){\circle*{3}}
\put(243,30){\circle*{3}}
\put(243,30){\vector(0,1){8}}
\put(245,38){\makebox(0,0)[l]{${\displaystyle '}$}}

\put(285,20){\framebox(10,10){}}
\put(285,30){\circle*{3}}
\put(295,30){\circle*{3}}
\put(295,30){\vector(0,-1){8}}
\put(297,28){\makebox(0,0)[l]{${\displaystyle '}$}}

\put(334,20){\line(1,0){10}}
\put(334,20){\line(0,1){8}}
\put(334,28){\line(5,2){10}}
\put(334,32){\line(0,1){8}}
\put(334,32){\line(5,-2){10}}
\put(334,40){\line(1,0){10}}
\put(344,20){\line(0,1){8}}
\put(344,32){\line(0,1){8}}
\put(344,32){\vector(0,1){6}}
\put(344,28){\vector(0,-1){6}}
\put(346,38){\makebox(0,0)[l]{${\displaystyle '}$}}

\put(364,30){\framebox(10,10){}}
\put(374,30){\vector(0,1){8}}
\put(376,38){\makebox(0,0)[l]{${\displaystyle '}$}}

\put(394,20){\framebox(10,10){}}
\put(404,30){\vector(0,-1){8}}
\put(406,28){\makebox(0,0)[l]{${\displaystyle '}$}}

\put(450,20){\framebox(10,9){}}
\put(450,31){\framebox(10,9){}}
\put(460,31){\vector(0,1){7}}
\put(460,29){\vector(0,-1){7}}
\put(462,28){\makebox(0,0)[l]{${\displaystyle '}$}}
\put(462,38){\makebox(0,0)[l]{${\displaystyle '}$}}

\end{picture}

\noindent
It is straightforward to show that, in general,
in the weak coupling, large $N$ limit, only shifted single
loops\footnote{For $SU(N)$, a single Wilson loop can
``wrap around'' itself up to $N-1$ times before it becomes
linearly dependent on clusters of simpler loops.}
survive in $S$ and $G$.
Why this appears to be a good approximation at $N=2$ remains unclear.

\pagebreak
{\Large \bf Conclusions}
\vspace{5pt}

In general,
the application of the coupled cluster method to lattice gauge
theories depends crucially on the rearrangement, or shifting, of
the clusters in $S$ and $G$ to describe gauge invariant fluctuations.
At $x \ll 1$ the method matches on to perturbation
theory\footnote{
The method, without shifting, in fact provides an efficient way,
avoiding any group integrals, of calculating
perturbative expansions for both eigenvalues and eigenvectors
due to the linked property of the Schr\"odinger equations.},
while at $x \gg 1$, it is always found that small, simple shifted clusters
dominate in the wavefunctions over large, complicated shifted clusters
due to combinatorical effects in the Schr\"odinger equations.
Furthermore, the interpretation of the discarded terms in the Schr\"odinger
equations as higher order fluctuations enables otherwise
formidable problems involving linear dependences among clusters to be largely
circumvented.

In summary, the shifted coupled cluster method provides
a non-perturbative semi-analytic approach to lattice gauge theories,
involving a direct physical approximation scheme, and gives
non-perturbative information
on both eigenvalues and eigenfunctions of the Schr\"odinger equations.
The results presented here show (table 1) that the method converges
very rapidly for the vacuum energy (assuming that the Pad\'e results
are a good guide) and should be capable of providing excellent
approximations to the vacuum wavefunction. Further work is needed
to establish how well it works for excited states;
through a more judicious choice of clusters in $S$ and $G$,
it should be possible to improve significantly upon
the results presented here for the mass gap at intermediate
couplings so that it can be reliably extrapolated to the weak coupling
limit.
The method therefore promises to provide much valuable analytic information
on the behaviour of lattice gauge theories. Fuller accounts of this
work will appear elsewhere \cite{inprep}.

\vspace{5pt}
{\Large \bf Acknowledgements}
\vspace{5pt}

N.J.W. acknowledges the financial support of an SERC/NATO Postdoctoral
Fellowship.

\pagebreak

\noindent
{\Large\bf Appendix: Fourth Order (Unshifted) Clusters}

\noindent
\begin{picture}(425,560)

\put(5,515){\makebox(0,0)[l]{${\displaystyle
C_{1,s} =
}$}}
\put(50,510){\framebox(10,10){}}

\put(5,455){\makebox(0,0)[l]{${\displaystyle
C_{2,s} =
}$}}
\put(50,450){\framebox(9,10){}}
\put(61,450){\framebox(9,10){}}

\put(5,425){\makebox(0,0)[l]{${\displaystyle
C_{3,s} =
}$}}
\put(50,420){\framebox(20,10){}}

\put(5,395){\makebox(0,0)[l]{${\displaystyle
C_{4,s} =
}$}}
\put(50,390){\framebox(10,10){}}
\put(52,392){\framebox(6,6){}}

\put(5,335){\makebox(0,0)[l]{${\displaystyle
C_{5,s} =
}$}}
\put(50,330){\framebox(9,9){}}
\put(61,330){\framebox(9,9){}}
\put(61,341){\framebox(9,9){}}

\put(5,305){\makebox(0,0)[l]{${\displaystyle
C_{6,s} =
}$}}
\put(50,300){\framebox(20,9){}}
\put(60,311){\framebox(10,9){}}

\put(5,275){\makebox(0,0)[l]{${\displaystyle
C_{7,s} =
}$}}
\put(50,270){\framebox(10,10){}}
\put(52,272){\framebox(6,6){}}
\put(62,270){\framebox(8,10){}}

\put(5,245){\makebox(0,0)[l]{${\displaystyle
C_{8,s} =
}$}}
\put(50,240){\framebox(20,10){}}
\put(52,242){\framebox(7,6){}}

\put(5,215){\makebox(0,0)[l]{${\displaystyle
C_{9,s} =
}$}}
\put(50,210){\framebox(9,10){}}
\put(61,210){\framebox(8,10){}}
\put(71,210){\framebox(9,10){}}

\put(5,185){\makebox(0,0)[l]{${\displaystyle
C_{10,s} =
}$}}
\put(50,180){\framebox(19,10){}}
\put(71,180){\framebox(9,10){}}

\put(5,155){\makebox(0,0)[l]{${\displaystyle
C_{11,s} =
}$}}
\put(50,150){\line(1,0){20}}
\put(50,150){\line(0,1){10}}
\put(50,160){\line(1,0){10}}
\put(60,160){\line(0,1){10}}
\put(60,170){\line(1,0){10}}
\put(70,150){\line(0,1){20}}

\put(5,125){\makebox(0,0)[l]{${\displaystyle
C_{12,s} =
}$}}
\put(50,120){\framebox(30,10){}}

\put(5,95){\makebox(0,0)[l]{${\displaystyle
C_{13,s} =
}$}}
\put(50,90){\framebox(10,10){}}
\put(52,92){\framebox(6,6){}}
\put(54,94){\framebox(2,2){}}

\put(5,35){\makebox(0,0)[l]{${\displaystyle
C_{14,s} =
}$}}
\put(50,30){\framebox(9,9){}}
\put(61,30){\framebox(8,9){}}
\put(71,30){\framebox(9,9){}}
\put(71,41){\framebox(9,9){}}

\put(5,5){\makebox(0,0)[l]{${\displaystyle
C_{15,s} =
}$}}
\put(50,0){\framebox(19,9){}}
\put(71,0){\framebox(9,9){}}
\put(71,11){\framebox(9,9){}}

\put(115,515){\makebox(0,0)[l]{${\displaystyle
C_{16,s} =
}$}}
\put(160,510){\framebox(9,9){}}
\put(162,512){\framebox(5,5){}}
\put(171,510){\framebox(9,9){}}
\put(171,521){\framebox(9,9){}}

\put(115,485){\makebox(0,0)[l]{${\displaystyle
C_{17,s} =
}$}}
\put(160,480){\framebox(9,9){}}
\put(171,480){\framebox(9,9){}}
\put(160,491){\framebox(9,9){}}
\put(171,491){\framebox(9,9){}}

\put(115,455){\makebox(0,0)[l]{${\displaystyle
C_{18,s} =
}$}}
\put(160,450){\framebox(9,9){}}
\put(171,450){\framebox(9,9){}}
\put(160,461){\framebox(20,9){}}

\put(115,425){\makebox(0,0)[l]{${\displaystyle
C_{19,s} =
}$}}
\put(160,420){\framebox(9,9){}}
\put(171,420){\framebox(9,9){}}
\put(173,422){\framebox(5,5){}}
\put(171,431){\framebox(9,9){}}

\put(115,395){\makebox(0,0)[l]{${\displaystyle
C_{20,s} =
}$}}
\put(160,390){\framebox(9,9){}}
\put(171,390){\framebox(9,20){}}
\put(173,392){\framebox(5,6){}}

\put(115,365){\makebox(0,0)[l]{${\displaystyle
C_{21,s} =
}$}}
\put(160,360){\framebox(9,9){}}
\put(171,360){\framebox(9,9){}}
\put(171,371){\framebox(9,9){}}
\put(182,371){\framebox(9,9){}}

\put(115,335){\makebox(0,0)[l]{${\displaystyle
C_{22,s} =
}$}}
\put(160,330){\framebox(9,9){}}
\put(171,330){\framebox(9,9){}}
\put(171,341){\framebox(20,9){}}

\put(115,305){\makebox(0,0)[l]{${\displaystyle
C_{23,s} =
}$}}
\put(160,300){\framebox(9,10){}}
\put(171,300){\framebox(9,20){}}
\put(173,312){\framebox(5,6){}}

\put(115,275){\makebox(0,0)[l]{${\displaystyle
C_{24,s} =
}$}}
\put(160,270){\framebox(9,9){}}
\put(171,281){\framebox(9,9){}}

\put(115,245){\makebox(0,0)[l]{${\displaystyle
C_{25,s} =
}$}}
\put(160,240){\framebox(9,9){}}
\put(171,240){\framebox(8,9){}}
\put(181,240){\framebox(9,9){}}
\put(171,251){\framebox(8,9){}}

\put(115,215){\makebox(0,0)[l]{${\displaystyle
C_{26,s} =
}$}}
\put(160,210){\framebox(19,9){}}
\put(181,210){\framebox(9,9){}}
\put(170,221){\framebox(9,9){}}

\put(115,185){\makebox(0,0)[l]{${\displaystyle
C_{27,s} =
}$}}
\put(160,180){\framebox(20,9){}}
\put(170,191){\framebox(10,9){}}
\put(172,193){\framebox(6,5){}}

\put(115,155){\makebox(0,0)[l]{${\displaystyle
C_{28,s} =
}$}}
\put(160,150){\line(1,0){20}}
\put(160,150){\line(0,1){10}}
\put(160,160){\line(1,0){10}}
\put(170,160){\line(0,1){10}}
\put(170,170){\line(1,0){10}}
\put(180,150){\line(0,1){20}}
\put(172,161){\framebox(6,7){}}

\put(115,125){\makebox(0,0)[l]{${\displaystyle
C_{29,s} =
}$}}
\put(160,120){\line(1,0){20}}
\put(160,120){\line(0,1){10}}
\put(160,130){\line(1,0){10}}
\put(170,130){\line(0,1){10}}
\put(170,140){\line(1,0){10}}
\put(180,120){\line(0,1){20}}
\put(160,132){\framebox(8,8){}}

\put(115,95){\makebox(0,0)[l]{${\displaystyle
C_{30,s} =
}$}}
\put(160,90){\framebox(9,9){}}
\put(171,90){\framebox(19,9){}}
\put(180,101){\framebox(10,9){}}

\put(115,65){\makebox(0,0)[l]{${\displaystyle
C_{31,s} =
}$}}
\put(160,60){\framebox(30,9){}}
\put(180,71){\framebox(10,9){}}

\put(115,35){\makebox(0,0)[l]{${\displaystyle
C_{32,s} =
}$}}
\put(160,30){\framebox(9,10){}}
\put(171,30){\framebox(8,20){}}
\put(181,40){\framebox(9,10){}}

\put(115,5){\makebox(0,0)[l]{${\displaystyle
C_{33,s} =
}$}}
\put(160,0){\line(1,0){20}}
\put(160,0){\line(0,1){10}}
\put(160,10){\line(1,0){10}}
\put(170,10){\line(0,1){10}}
\put(170,20){\line(1,0){10}}
\put(180,0){\line(0,1){20}}
\put(182,10){\framebox(8,10){}}

\put(225,515){\makebox(0,0)[l]{${\displaystyle
C_{34,s} =
}$}}
\put(270,510){\framebox(9,10){}}
\put(281,510){\framebox(8,20){}}
\put(291,510){\framebox(9,10){}}

\put(225,485){\makebox(0,0)[l]{${\displaystyle
C_{35,s} =
}$}}
\put(270,480){\line(1,0){20}}
\put(270,480){\line(0,1){10}}
\put(270,490){\line(1,0){10}}
\put(280,490){\line(0,1){10}}
\put(280,500){\line(1,0){10}}
\put(290,480){\line(0,1){20}}
\put(292,480){\framebox(8,10){}}

\put(225,455){\makebox(0,0)[l]{${\displaystyle
C_{36,s} =
}$}}
\put(270,450){\framebox(30,9){}}
\put(280,461){\framebox(10,9){}}

\put(225,425){\makebox(0,0)[l]{${\displaystyle
C_{37,s} =
}$}}
\put(270,420){\framebox(9,10){}}
\put(281,420){\framebox(8,10){}}
\put(291,420){\framebox(9,20){}}

\put(225,395){\makebox(0,0)[l]{${\displaystyle
C_{38,s} =
}$}}
\put(270,390){\framebox(19,10){}}
\put(291,390){\framebox(9,20){}}

\put(225,365){\makebox(0,0)[l]{${\displaystyle
C_{39,s} =
}$}}
\put(270,360){\framebox(20,9){}}
\put(270,371){\framebox(20,9){}}

\put(225,335){\makebox(0,0)[l]{${\displaystyle
C_{40,s} =
}$}}
\put(270,330){\framebox(20,9){}}
\put(280,341){\framebox(20,9){}}

\put(225,305){\makebox(0,0)[l]{${\displaystyle
C_{41,s} =
}$}}
\put(270,300){\framebox(9,10){}}
\put(272,302){\framebox(5,6){}}
\put(281,300){\framebox(8,10){}}
\put(291,300){\framebox(9,10){}}

\put(225,275){\makebox(0,0)[l]{${\displaystyle
C_{42,s} =
}$}}
\put(270,270){\framebox(9,10){}}
\put(272,272){\framebox(5,6){}}
\put(281,270){\framebox(19,10){}}

\put(225,245){\makebox(0,0)[l]{${\displaystyle
C_{43,s} =
}$}}
\put(270,240){\framebox(9,10){}}
\put(272,242){\framebox(5,6){}}
\put(281,240){\framebox(9,10){}}
\put(283,242){\framebox(5,6){}}

\put(225,215){\makebox(0,0)[l]{${\displaystyle
C_{44,s} =
}$}}
\put(270,210){\framebox(10,10){}}
\put(272,212){\framebox(6,6){}}
\put(274,214){\framebox(2,2){}}
\put(282,210){\framebox(8,10){}}

\put(225,185){\makebox(0,0)[l]{${\displaystyle
C_{45,s} =
}$}}
\put(270,180){\framebox(20,10){}}
\put(272,182){\framebox(7,6){}}
\put(274,184){\framebox(3,2){}}

\put(225,155){\makebox(0,0)[l]{${\displaystyle
C_{46,s} =
}$}}
\put(270,150){\framebox(20,10){}}
\put(272,152){\framebox(7,6){}}
\put(281,152){\framebox(7,6){}}

\put(225,125){\makebox(0,0)[l]{${\displaystyle
C_{47,s} =
}$}}
\put(270,120){\framebox(8,10){}}
\put(280,120){\framebox(10,10){}}
\put(282,122){\framebox(6,6){}}
\put(292,120){\framebox(8,10){}}

\put(225,95){\makebox(0,0)[l]{${\displaystyle
C_{48,s} =
}$}}
\put(270,90){\framebox(19,10){}}
\put(281,92){\framebox(6,6){}}
\put(291,90){\framebox(9,10){}}

\put(225,65){\makebox(0,0)[l]{${\displaystyle
C_{49,s} =
}$}}
\put(270,60){\framebox(30,10){}}
\put(281,62){\framebox(8,6){}}

\put(225,35){\makebox(0,0)[l]{${\displaystyle
C_{50,s} =
}$}}
\put(270,30){\line(1,0){20}}
\put(270,30){\line(0,1){10}}
\put(270,40){\line(1,0){10}}
\put(280,40){\line(0,1){10}}
\put(280,50){\line(1,0){10}}
\put(290,30){\line(0,1){20}}
\put(282,32){\framebox(6,6){}}

\put(225,5){\makebox(0,0)[l]{${\displaystyle
C_{51,s} =
}$}}
\put(270,0){\framebox(19,10){}}
\put(272,2){\framebox(7,6){}}
\put(291,0){\framebox(9,10){}}

\put(335,515){\makebox(0,0)[l]{${\displaystyle
C_{52,s} =
}$}}
\put(380,510){\framebox(30,10){}}
\put(382,512){\framebox(7,6){}}

\put(335,485){\makebox(0,0)[l]{${\displaystyle
C_{53,s} =
}$}}
\put(380,480){\framebox(20,10){}}
\put(382,482){\framebox(16,6){}}

\put(335,455){\makebox(0,0)[l]{${\displaystyle
C_{54,s} =
}$}}
\put(380,450){\framebox(9,10){}}
\put(391,450){\framebox(8,10){}}
\put(401,450){\framebox(8,10){}}
\put(411,450){\framebox(9,10){}}

\put(335,425){\makebox(0,0)[l]{${\displaystyle
C_{55,s} =
}$}}
\put(380,420){\framebox(19,10){}}
\put(401,420){\framebox(8,10){}}
\put(411,420){\framebox(9,10){}}

\put(335,395){\makebox(0,0)[l]{${\displaystyle
C_{56,s} =
}$}}
\put(380,390){\framebox(10,10){}}
\put(400,390){\framebox(10,10){}}

\put(335,365){\makebox(0,0)[l]{${\displaystyle
C_{57,s} =
}$}}
\put(380,360){\framebox(9,10){}}
\put(391,360){\framebox(18,10){}}
\put(411,360){\framebox(9,10){}}

\put(335,335){\makebox(0,0)[l]{${\displaystyle
C_{58,s} =
}$}}
\put(380,330){\framebox(29,10){}}
\put(411,330){\framebox(9,10){}}

\put(335,305){\makebox(0,0)[l]{${\displaystyle
C_{59,s} =
}$}}
\put(390,300){\line(1,0){20}}
\put(390,300){\line(0,1){10}}
\put(390,310){\line(1,0){10}}
\put(400,310){\line(0,1){10}}
\put(400,320){\line(1,0){10}}
\put(410,300){\line(0,1){20}}
\put(380,300){\framebox(8,10){}}

\put(335,275){\makebox(0,0)[l]{${\displaystyle
C_{60,s} =
}$}}
\put(380,270){\framebox(19,10){}}
\put(401,270){\framebox(19,10){}}

\put(335,245){\makebox(0,0)[l]{${\displaystyle
C_{61,s} =
}$}}
\put(380,240){\line(1,0){30}}
\put(380,240){\line(0,1){10}}
\put(380,250){\line(1,0){20}}
\put(400,250){\line(0,1){10}}
\put(400,260){\line(1,0){10}}
\put(410,240){\line(0,1){20}}

\put(335,215){\makebox(0,0)[l]{${\displaystyle
C_{62,s} =
}$}}
\put(380,210){\framebox(20,20){}}

\put(335,185){\makebox(0,0)[l]{${\displaystyle
C_{63,s} =
}$}}
\put(380,180){\line(1,0){20}}
\put(380,180){\line(0,1){10}}
\put(380,190){\line(1,0){10}}
\put(390,190){\line(0,1){10}}
\put(390,200){\line(1,0){20}}
\put(400,180){\line(0,1){10}}
\put(400,190){\line(1,0){10}}
\put(410,190){\line(0,1){10}}

\put(335,155){\makebox(0,0)[l]{${\displaystyle
C_{64,s} =
}$}}
\put(380,150){\line(1,0){30}}
\put(380,150){\line(0,1){10}}
\put(380,160){\line(1,0){10}}
\put(390,160){\line(0,1){10}}
\put(390,170){\line(1,0){10}}
\put(400,160){\line(0,1){10}}
\put(400,160){\line(1,0){10}}
\put(410,150){\line(0,1){10}}

\put(335,125){\makebox(0,0)[l]{${\displaystyle
C_{65,s} =
}$}}
\put(380,120){\line(1,0){11}}
\put(380,120){\line(0,1){11}}
\put(380,131){\line(1,0){9}}
\put(389,131){\line(0,1){9}}
\put(389,140){\line(1,0){11}}
\put(391,120){\line(0,1){9}}
\put(391,129){\line(1,0){9}}
\put(400,129){\line(0,1){11}}

\put(335,95){\makebox(0,0)[l]{${\displaystyle
C_{66,s} =
}$}}
\put(380,90){\framebox(40,10){}}

\put(335,65){\makebox(0,0)[l]{${\displaystyle
C_{67,s} =
}$}}
\put(380,60){\line(1,0){30}}
\put(380,60){\line(0,1){10}}
\put(380,70){\line(1,0){10}}
\put(390,62){\line(0,1){8}}
\put(390,62){\line(1,0){10}}
\put(400,62){\line(0,1){8}}
\put(400,70){\line(1,0){10}}
\put(410,60){\line(0,1){10}}

\put(335,35){\makebox(0,0)[l]{${\displaystyle
C_{68,s} =
}$}}
\put(380,30){\line(1,0){20}}
\put(380,30){\line(0,1){9}}
\put(380,39){\line(1,0){9}}
\put(389,32){\line(0,1){7}}
\put(389,32){\line(1,0){9}}
\put(398,32){\line(0,1){9}}
\put(391,41){\line(1,0){7}}
\put(391,41){\line(0,1){9}}
\put(391,50){\line(1,0){9}}
\put(400,30){\line(0,1){20}}

\put(335,5){\makebox(0,0)[l]{${\displaystyle
C_{69,s} =
}$}}
\put(380,0){\framebox(10,10){}}
\put(381.5,1.5){\framebox(7,7){}}
\put(383,3){\framebox(4,4){}}
\put(384.5,4.5){\framebox(1,1){}}

\end{picture}

\pagebreak

\pagebreak
{\Large\bf Figure Captions}
\vspace{10pt}

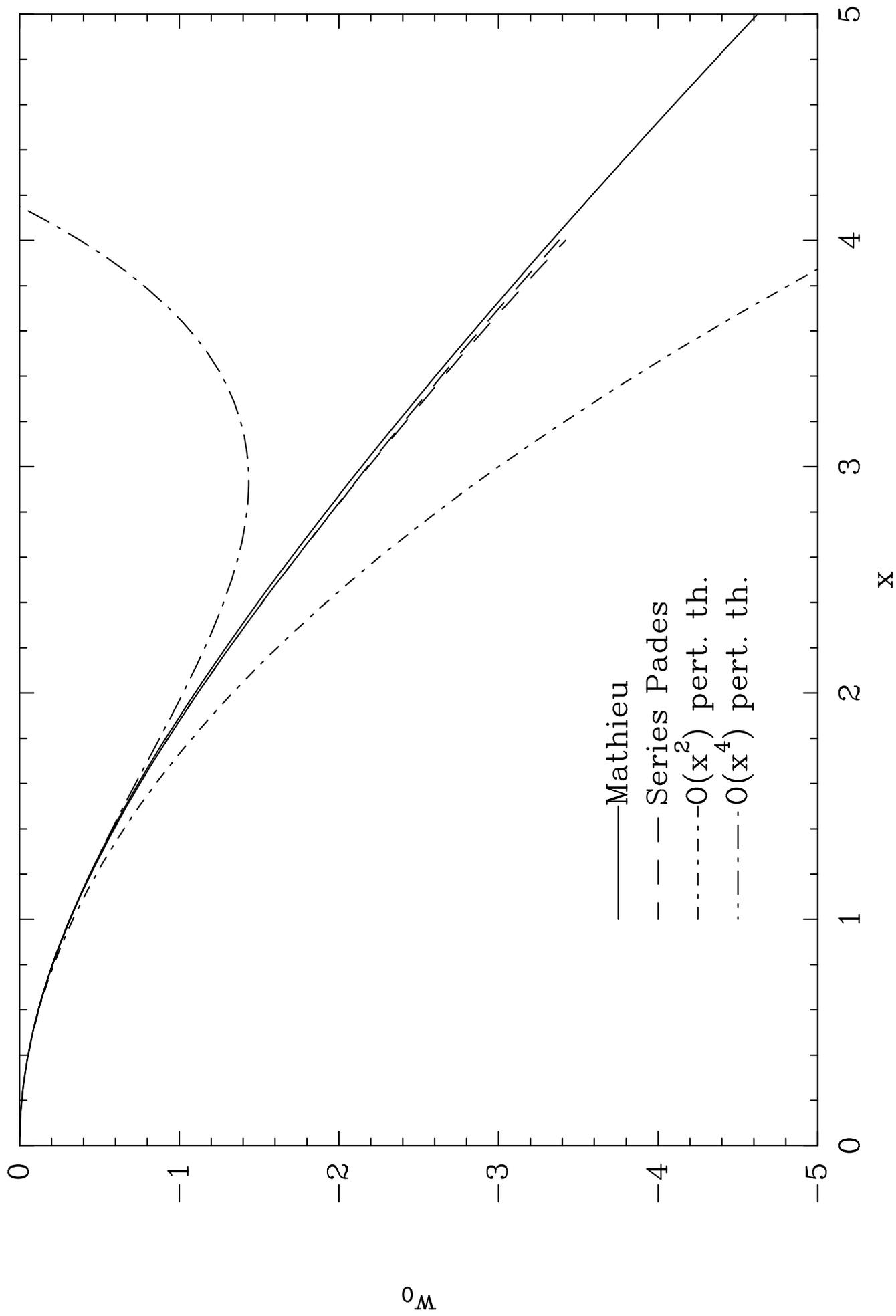
Fig. 1. Graph of the vacuum energy $w_{0}$ vs. $x$ for the exact
Mathieu solution of the one-plaquette-cutoff coupled cluster equation (13),
together with
i) the Pad\'e approximants
derived from the ${\cal O}(x^{14})$ strong coupling perturbation
theory series
taken from \cite{hamirv3d}, and
ii) the results
from ${\cal O}(x^{2})$ and ${\cal O}(x^{4})$
strong coupling perturbation theory, which show that the Mathieu
result is non-trivial.

\vspace{10pt}

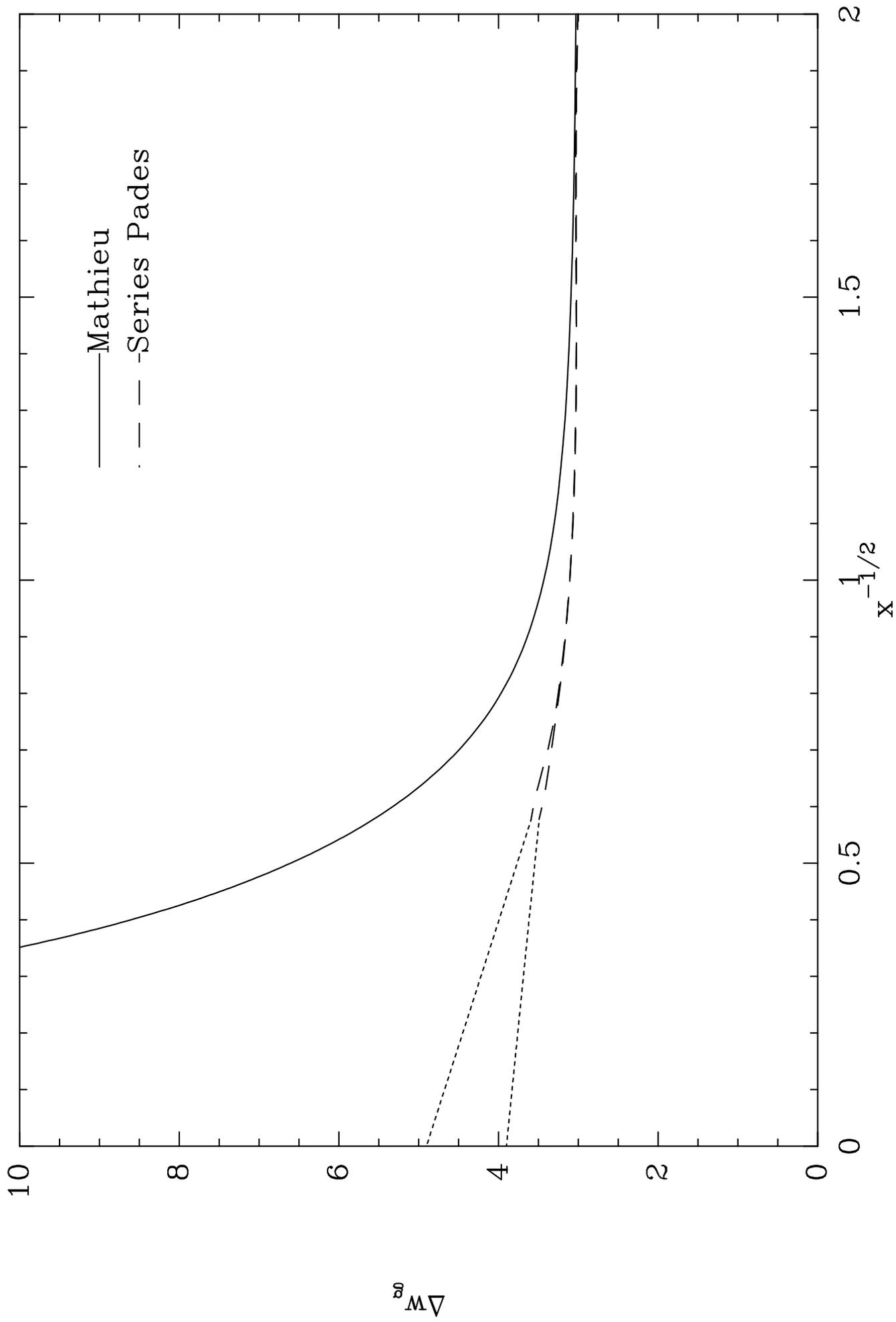
Fig. 2. Graph of the mass gap $\Delta w_{g}$ vs. $x^{-1/2}$ for the exact
Mathieu solution of the one-plaquette-cutoff coupled cluster equation,
together with
i) the Pad\'e approximants
derived from the ${\cal O}(x^{10})$ strong coupling perturbation
theory series
taken from \cite{hamirv3d}, and
ii) the estimated bounds, shown dotted,
for their extrapolation to the continuum limit $x \rightarrow\infty$, also
from \cite{hamirv3d}.

\vspace{10pt}

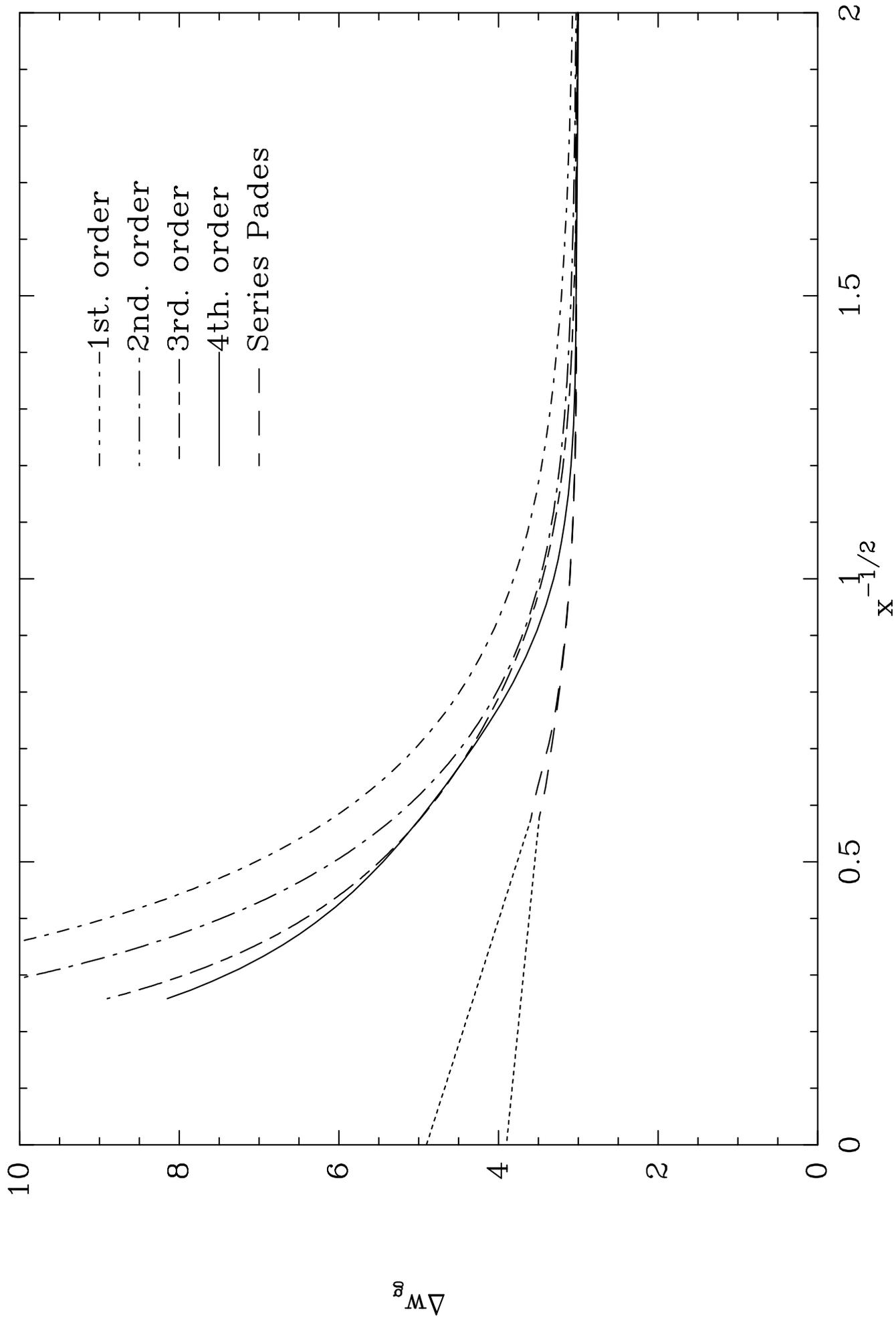
Fig. 3. Graph of the mass gap $\Delta w_{g}$ vs. $x^{-1/2}$ for
the shifted coupled cluster
wavefunctions $\Psi_{g} = G\,{\rm exp}S$ with $S$, $G$ including
the sets of clusters which, unshifted, occur at the first four orders
of strong coupling perturbation theory. Also shown are
the Pad\'e approximants and continuum extrapolations
derived from the ${\cal O}(x^{10})$ strong coupling perturbation
theory series
taken from \cite{hamirv3d}.

\vspace{10pt}

Fig. 4. Graph of the absolute values $|a_{i}'/a_{1}'|$ vs. $x$
for the amplitudes $a_{i}'$ of the most dominant of the 69 shifted
clusters occuring in the vacuum wavefunction $\Psi_{0} = {\rm exp}S$ at
the 4th order approximation, where $a_{1}'$ is the amplitude of
the shifted single plaquette loop $C_{1}'$.
The cluster labels $i$ are shown on the right side of the figure.

\vspace{10pt}

Fig. 5. Graph of the ampitudes $b_{i}'$ vs. $x$
for the most dominant of the 69 shifted
clusters occuring in the scalar
glueball wavefunction $\Psi_{g} = G\,{\rm exp}S$ at
the 4th order approximation.
The cluster labels $i$ are shown on the right side of the figure.

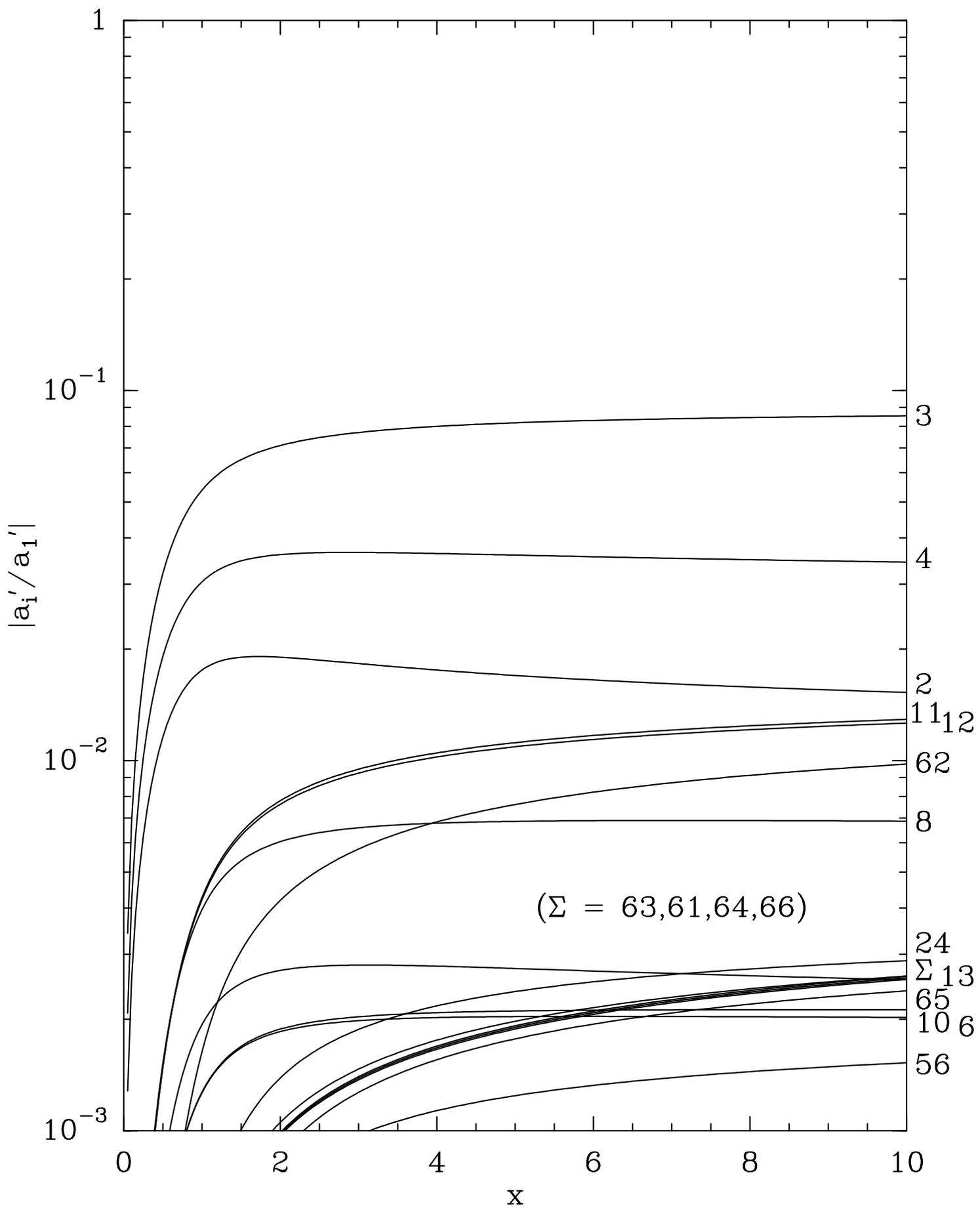
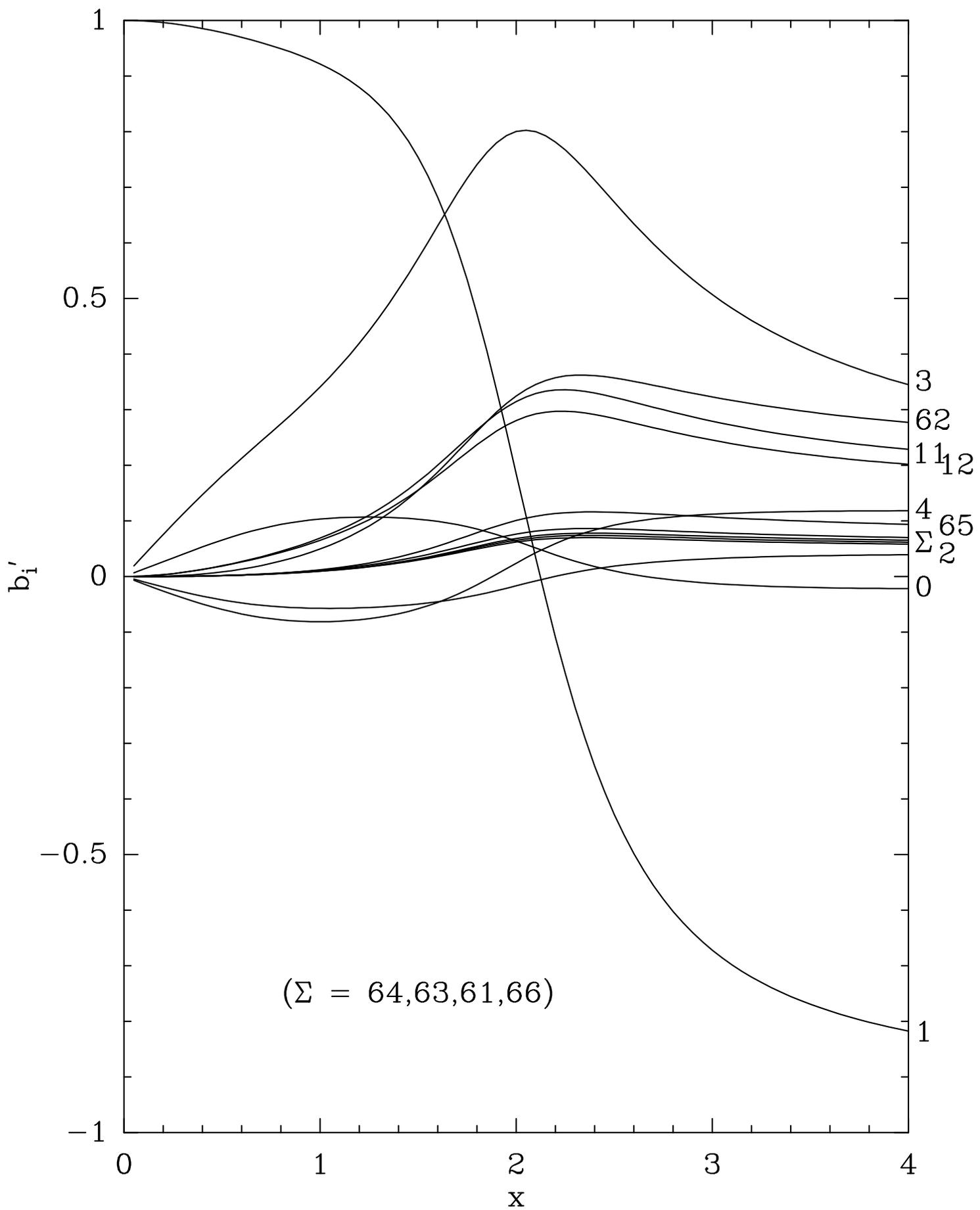
\end{document}